\documentclass[12pt,prd,nofootinbib,preprintnumbers]{revtex4}
\usepackage{graphicx,amsmath,amssymb}

\usepackage{epsfig}

\newcommand{\be}{\begin{equation}}
\newcommand{\ee}{\end{equation}}
\newcommand{\bea}{\begin{eqnarray}}
\newcommand{\eea}{\end{eqnarray}}

\def\le{\left}
\def\ri{\right}

\begin{document}

\title {Adventures in Holographic Dimer Models}

\preprint{SU-ITP-10/26, SLAC-PUB-14252}

\author{Shamit Kachru}
\affiliation{Department of Physics and SLAC, Stanford University, Stanford, CA 94305, USA\\
{\tt skachru@stanford.edu}}

\author{Andreas Karch}
\affiliation{ Department of Physics, University of Washington, Seattle, WA 98195-1560, USA\\
{\tt karch@phys.washington.edu}}

\author{Sho Yaida}
\affiliation{Department of Physics, Stanford University, Stanford, CA 94305, USA\\
{\tt yaida@stanford.edu}}


\begin{abstract}
We abstract the essential features of holographic dimer models, and develop several new applications of these models.
First, semi-holographically coupling free band fermions to holographic dimers, we uncover novel phase transitions between conventional Fermi liquids and non-Fermi liquids, accompanied by a change in the structure of the Fermi surface. Second, we make dimer vibrations propagate through the whole crystal by way of double trace deformations, obtaining nontrivial band structure. In a simple toy model, the topology of the band structure experiences an interesting reorganization as we vary the strength of the double trace deformations. Finally, we develop tools that would allow one to build, in a bottom-up fashion, a holographic avatar of the Hubbard model.
\end{abstract}

\maketitle

\section{Introduction}

Holographic models of condensed matter systems have seen a recent surge in interest. While this program can by now point to a few successes, there is one feature of realistic solids that is commonly not shared by their holographic stand-ins: in real solids, translation invariance is broken to a discrete subgroup by formation of a lattice.\footnote{A second, somewhat related, feature which is lacking in most holographic studies of condensed matter systems is disorder. For early studies of how to include disorder, see \cite{Hartnoll:2008hs,Fujita:2008rs}.} In the holographic models with translational symmetry unbroken, momentum is strictly conserved as there are no Umklapp processes to dissipate it. Consequently, in the background of an electric field, energy and (at finite density) momentum are pumped into the system at a constant rate. This leads to unrealistic transport properties such as an unsmeared delta-function Drude peak, among other things.\footnote{One way to work around this is to make use of the large number of internal degrees of freedom, which is one of the defining features of holographic models. A plasma sea with order $N^2$ excitations, where $N$ is a large integer, can effectively act as a heat and momentum dump for charge carriers, leading to interesting DC conductivities \cite{Karch:2007pd,Hartnoll:2008vx,Faulkner:2010da}.} Also, in heavy fermion materials, strong correlations of itinerant electrons with localized spins on a lattice are believed to trigger interesting phenomena such as quantum criticality. It is desirable to find holographic models which exhibit lattice structure.

In~\cite{Kachru:2009xf}, we constructed holographic models with translational symmetry explicitly broken by fermionic degrees of freedom localized on lattice sites, interacting with a continuum gauge field.\footnote{An alternative proposal in which the whole gauge theory was forced to live on a lattice was put forward in \cite{Hellerman:2002qa}.
A method for incorporating lattice defect fermions through semi-holographic techniques, which works only when the defect fermions are ${\it neutral}$ under the large $N$
gauge group, appears in \cite{Faulkner:2010tq}.  Other lattice defect models with related features to the models we study will appear in \cite{Eva}.
} These systems have many appealing features. They naturally give rise to bulk fermions living on a lattice of AdS$_2$ spacetimes; such spacetimes play a crucial role in holographic non-Fermi liquids \cite{Lee:2008xf,Liu:2009dm,Faulkner:2009wj,Cubrovic:2009ye}. Thus they provide a natural home to study holographic non-Fermi liquids without worrying about the physics of the asymptotically anti-de Sitter (AdS) Reissner-Nordstr\"{o}m black brane, especially its large ground state degeneracy and potential instabilities. It has also been argued from the field theory point of view \cite{Sachdev:2010um} that a lattice of localized defect fermions interacting with continuum degrees of freedom is the best candidate for a condensed matter system exhibiting the phenomenology of holographic non-Fermi liquids

In this paper, we abstract a few essential features of the holographic dimer models constructed in \cite{Kachru:2009xf}, and then extend our knowledge of these models in several directions. First, we add free band fermions to the boundary theory and then weakly mix them with the large $N$ sector semi-holographically, following Faulkner and Polchinski~\cite{Faulkner:2010tq}. When the large $N$ sector undergoes a dimerization transition, it induces a dramatic change in the singlet fermion sector. Namely, the melting of dimers turns a normal Fermi liquid into a non-Fermi liquid, with an accompanying change in the structure of its Fermi surface. This transition may be of some interest in relation to toy models of heavy fermion materials~\cite{Sachdev:2010um}.

We then turn to a study of dimer vibrations. In the original holographic dimer models, the dimers do not effectively talk to each other in the large $N$ limit, and each has a discrete vibration spectrum. By adding double trace deformations, we let the dimers communicate with each other, allowing their vibrations to propagate through the whole crystal in the form of  Bloch waves.  As an illustration, we explicitly work out the band structure in a solvable toy model of this type.  Interestingly, we observe a reorganization in the topology of the band structure as we vary the strength of the double trace deformations.

Finally, we also outline how one would deform the system by an actual fermion hopping term so as to liberate otherwise immobile localized defect fermions. The corresponding problem in the bulk is beyond current brane technology, as one would have to know the non-Abelian generalization of the Dirac-Born-Infeld (DBI) action describing the fluctuations of the brane. Only the lowest few terms in a derivative expansion of this action are currently understood. However the deformation we identify can easily be implemented in a bottom-up model, where the bulk theory is not taken to be a system realized in string theory but instead is governed by an effective action with free parameters which can be matched against ``experiments." With this toolkit it is in principle possible to construct holographic avatars of one's favorite lattice models such as the Hubbard model. This should allow one to parametrize his/her ignorance in terms of the coefficients in the bulk effective action. While such a program has been somewhat successful for QCD \cite{Erlich:2005qh,DaRold:2005zs}, we will refrain from any such attempts in this work.

The organization of this paper is as follows.  We first abstract out the salient aspects of our previous construction~\cite{Kachru:2009xf} in Sec.~\ref{essence}.  In Sec.~\ref{semi} we then describe semi-holographic phase transitions between Fermi liquids and non-Fermi liquids, much in the spirit of
\cite{Faulkner:2010tq}. Changing gears, in Sec.~\ref{band}, we show how one can map out the band structure of dimer vibrations in the holographic dimer models perturbed by simple double trace deformations. For a soluble toy model, we observe reorganization in the topology of the band structure as the strength of the double trace deformations increases. Lastly, in Sec.~\ref{NAhopping}, we outline how knowledge of the non-Abelian DBI action would enable us to study a genuine fermion hopping deformation and how one could go about building holographic bottom-up models of generic strongly coupled lattices.  Detailed calculations supporting the plots of band structure in Sec. \ref{band}
have been relegated to the appendix, ``Hypergeometric-ology."

\section{Physics of a holographic dimer}
\label{essence}

\begin{figure}[t]
\includegraphics[scale=0.65,angle=0]{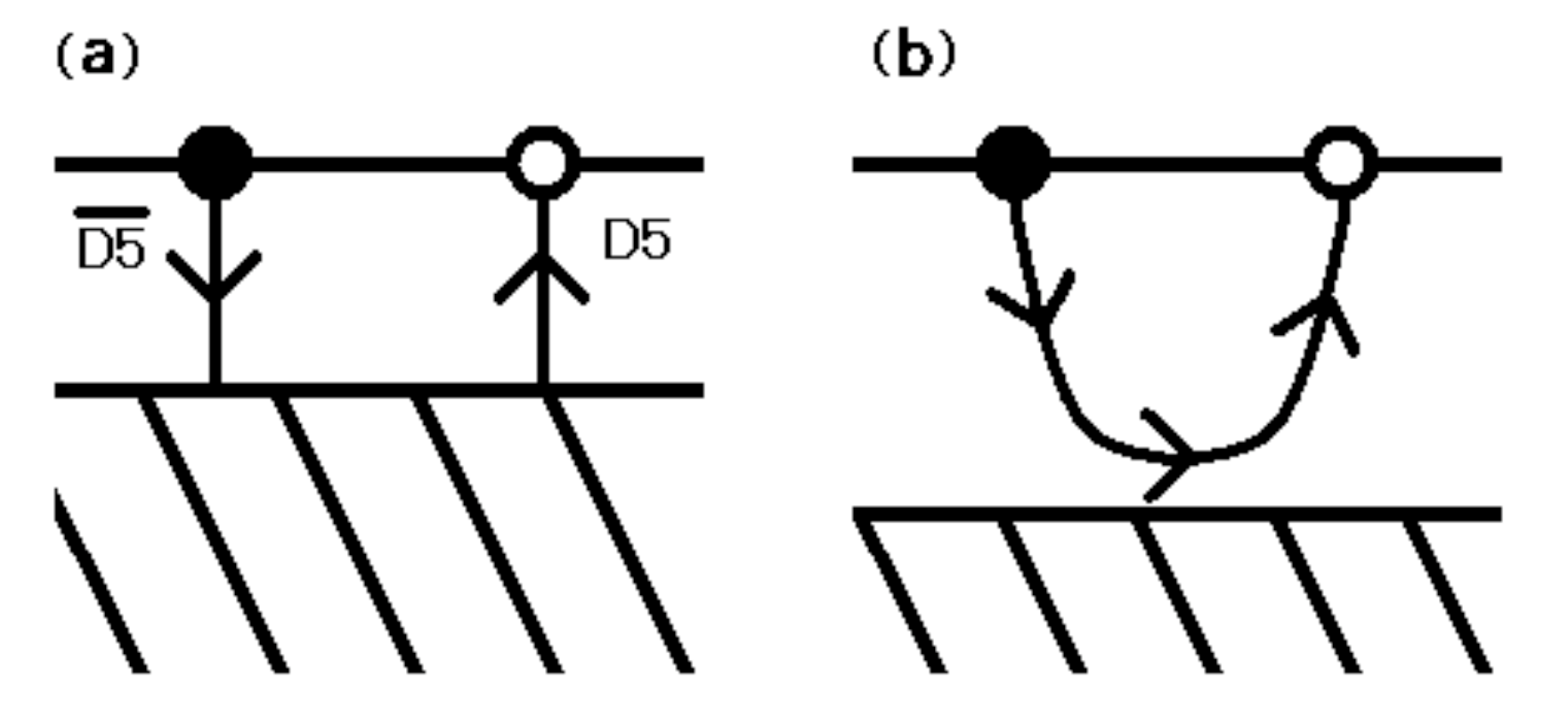}
\caption{(a)The high-temperature phase with $U(1)\times U(1)$ symmetry.  (b)The low-temperature phase with $U(1) \times U(1) \to U(1)$ via brane recombination.}
 \label{1reconnection}
\end{figure}

We first review the essential picture of a single dimer in the holographic dimer lattice models constructed in~\cite{Kachru:2009xf}. We began with a pair consisting of a D5-brane and an anti-D5-brane, with the bulk spacetime being the ${\rm AdS}_5$ Schwarzschild black brane [see Fig.1(a)]. Asymptotically, they wrap copies of ${\rm AdS}_2\times S^4$ in
the  ${\rm AdS}_5\times S^5$. In the boundary field theory, they introduce localized defect fermions, transforming under fundamental and antifundamental representations of the $SU(N)$ gauge group of the continuum gauge field, respectively. As we cool down the system, the size of the black brane shrinks down, and the D5-brane and anti-D5-brane pair up [see Fig.1(b)], spontaneously breaking $U(1)\times U(1)$ down to a diagonal $U(1)$. In the boundary field theory, this transition is characterized by a mildly nonlocal order parameter involving fermions from neighboring sites as well as an open Wilson line to insure gauge invariance \cite{Aharony:2008an}. This is the essence of the dimerization transition worked out in detail in~\cite{Kachru:2009xf}.

We now look, rather abstractly, at fluctuations of these probe branes.

\subsection{Undimerized phase: gapless spectrum}
In this phase, we can study each probe brane separately. For every possible fluctuation of the probe brane, there exists a corresponding gauge singlet operator localized at a point, ${\cal O}_J$, bosonic or fermionic. Here $J$ labels whether the operator is associated with a D5-brane or anti-D5-brane, and it will be promoted to a lattice index when we discuss a lattice of dimers.

By studying fluctuations in the bulk, it is in principle possible to work out
\be
\int dt e^{i\omega t}\langle {\cal O}_J(t) {\cal O}_{J'}^{\dagger}(0)\rangle=i \delta_{J, J'} {\cal G}(\omega)
\ee
in detail. For example, had we stayed in the undimerized phase down to zero temperature, the induced metric on the (anti-)D5-brane is exactly that of AdS$_2$ and $(0+1)$-dimensional conformality would dictate
\be
{\cal G}(\omega)=c \omega^{2\Delta -1}
\ee
with $c$ a calculable complex number and $\Delta$ the operator dimension of ${\cal O}_{J}$.\footnote{Strictly speaking, we can only trust our analysis in the undimerized phase down to temperatures of order $T^3 a^3 \sim \sqrt{\lambda}/N$. At this point the backreaction of the D5-branes on the background geometry can no longer be neglected \cite{Kachru:2009xf}. For most of this work we can safely neglect this complication as we are shielded from these parametrically small temperatures by the dimerization phase transition which occurs at $Ta \sim 1$. However, if we considered a lattice made purely of D5-branes instead of alternating D5- and anti-D5-branes
(as was also discussed in~\cite{Kachru:2009xf}), there would be no dimerization transition. In that case the physics would be well
captured by the AdS$_2$ gravity background down to this parametrically very low temperature.}

When we turn to semi-holographic constructions, it is crucial to keep in mind that ${\cal G}(\omega)$ in our concrete holographic lattice model behaves differently from
the Green's function for ${\rm AdS}_2\times {\bf R}^{d-1}$ at finite temperature, appearing in the construction of holographic non-Fermi liquids. This is because the metric on the embedded D5-brane, induced from the ${\rm AdS}_5$ Schwarzschild black brane metric, is different from the near-horizon geometry of the AdS Reissner-Nordstr\"{o}m black brane.

Nevertheless, one generic behavior of this type of model is that, since the probe brane is touching the horizon in this phase, the spectrum is gapless. This means that $\lim_{\omega\rightarrow0}{\cal G}(\omega)=0$.

\subsection{Dimerized phase: gapped spectrum}

In this phase, probe branes recombine and no longer stretch down to the horizon, leading to a gapped spectrum. Still, the induced metric near the asymptotic boundary is that of ${\rm AdS}_2$. Let us write the asymptotic AdS$_2$ metric as
\be
\label{nearboundarymetric}
ds^2_{{\rm AdS}_2} = \frac{1}{z_J^2}( - dt^2 + dz_J^2).
\ee
Then near the boundary point where the anti-D5-brane stretches down ($z_{\overline {\rm D5}}=0$), a time-dependent fluctuation with frequency $\omega$ can be expanded as
\be
e^{-i\omega t}\left[\alpha_{\overline {\rm D5}}(\omega) \left\{z_{\overline {\rm D5}}^{1-\Delta}+...\right\}+ \beta_{\overline {\rm D5}}(\omega) \left\{z_{\overline {\rm D5}}^{\Delta}+...\right\}\right]
\ee
whereas near the boundary point where the connected D5-brane stretches up ($z_{{\rm D5}}=0$), the same fluctuation can be expanded as
\be
e^{-i\omega t}\left[\alpha_{{\rm D5}}(\omega) \left\{z_{{\rm D5}}^{1-\Delta}+...\right\} + \beta_{{\rm D5}}(\omega)  \left\{z_{{\rm D5}}^{\Delta}+...\right\}\right].
\ee
The coefficients $\alpha_{\overline {\rm D5}}(\omega)$, $\beta_{\overline {\rm D5}}(\omega)$, $\alpha_{{\rm D5}}(\omega)$, and $\beta_{{\rm D5}}(\omega)$ are related by a frequency-dependent matrix as follows:
\be
\le(\begin{array}{ c c }
\alpha_{\rm D5}(\omega) \\
\beta_{\rm D5}(\omega) \end{array} \ri)
=
\le(\begin{array}{ c c }
t_{11}(\omega) & t_{12}(\omega) \\
t_{21}(\omega) & t_{22}(\omega)\end{array} \ri)
\le(\begin{array}{ c c }
\alpha_{\overline{\rm D5}}(\omega) \\
\beta_{\overline{\rm D5}}(\omega)\end{array} \ri)
\equiv {\mathbb T}(\omega) \, \le(\begin{array}{ c c }
\alpha_{\overline{\rm D5}}(\omega) \\
\beta_{\overline{\rm D5}}(\omega)\end{array} \ri).
\ee
For a simple phenomenological toy model, ${\mathbb T}(\omega)$ is worked out in detail in Appendix~\ref{hyper}.

In the absence of any deformation of the theory, there exists a steady dimer vibration with frequency $\omega_n$ if and only if there exists a consistent nontrivial solution with $\alpha_{\overline{\rm D5}}(\omega_n)=\alpha_{{\rm D5}}(\omega_n)=0$. This can happen if and only if $t_{12}(\omega_n)=0$. This gives rise to a pole in $\int dt e^{i\omega t}\langle {\cal O}_J(t) {\cal O}_{J'}^{\dagger}(0)\rangle$ at $\omega=\omega_n$.

Typically there exists a gap to the first dimer excitation and at low frequency correlation functions give
\be
\lim_{\omega\rightarrow0}\int dt e^{i\omega t}\langle {\cal O}_J(t) {\cal O}_{J'}^{\dagger}(0)\rangle=iA_{J, J'},
\ee
with all the components of $A_{J, J'}$ generically nonzero.

\

In passing, we note that for the top-down D3/D5 system of \cite{Kachru:2009xf}, the worldvolume gauge field and the slipping mode scalar mix due to the Wess-Zumino terms in the action. The mixed sector gives rise to two towers of scalar fields depending on the angular momentum $l$ on the internal sphere. As shown in \cite{Camino:2001at} they behave like fields with $m_l^2 = (l+3)(l+4)$ and $m_l^2=l (l-1)$, respectively. At $l=0$ we are hence effectively describing a massless scalar, presumably dual to the
defect fermion bilinear, as well as a massive scalar mode dual to a dimension 4 operator.

\

Now that we have gathered essential information regarding the dimerization transition and spectra in both phases, let us look at several physical applications. Anticipating the huge landscape of large $N$ dimer models, we will keep $\Delta$ and all the other information (computable in explicit models) as undetermined free parameters. From here on, each subsequent section of the paper can be read independently.

\section{Semi-holographic phase transitions}
\label{semi}

A rich set of non-Fermi liquid behaviors has recently been discovered by studying the physics of probe fermions in the asymptotically AdS Reissner-Nordstr\"{o}m background
\cite{Lee:2008xf,Liu:2009dm,Faulkner:2009wj,Cubrovic:2009ye}.
The near-horizon ${\rm AdS}_2 \times {\bf R}^{2}$ region of the black brane plays a crucial role in organizing and explaining this physics; the physics of the emergent ``locally quantum critical" theory dual to the AdS$_2$ region is what gives rise to the non-Fermi liquid behavior. However, this black brane is in some ways nongeneric.  For instance, it suffers from
 a superconducting instability in the presence of generic charged scalar fields in the bulk \cite{Gubser:2008px}, and neutral scalar fields coupled to the
 kinetic term of the bulk $U(1)$ gauge field deform the near-horizon geometry \cite{Goldstein:2009cv} to be of the Lifshitz form \cite{Kachru:2008yh}.  Even the backreaction
 of the fermions themselves deforms the near-horizon geometry to Lifshitz form at subleading orders in $1/N$  \cite{Hartnoll:2009ns} [shifting the ${\rm AdS}_2 \times {\bf R}^2$ geometry, which has
 a dynamical critical exponent $z = \infty$, to instead have $z \sim N$].  While in many cases these deformations may
 leave the essential physics of the fermion spectral function unchanged (see \cite{Faulkner:2010tq} for a nice discussion), it is also reasonable to find
 other ways that the essential insights of \cite{Lee:2008xf,Liu:2009dm,Faulkner:2009wj,Cubrovic:2009ye} can be reproduced in a more robust setting. The AdS$_2$ regions spanned by the D5- and anti-D5-branes in the top-down holographic dimer model of \cite{Kachru:2009xf} provide an alternative way to obtain the same physics. Here, we explore this in a semi-holographic setting following \cite{Faulkner:2010tq}, and we abstract the main features of the top-down model to include more generic possibilities.

We begin with a large $N$ field theory, governed by some action $S_{\rm strong}$, with the following features:
 \begin{enumerate}
 \item There is a lattice of defect fermions which undergoes a dimerization transition as we vary the external parameters such as temperature [see Fig.2 for the (1+1)-dimensional case].  We will focus on the cases for which this parameter is temperature, but one can easily generalize.\footnote{For instance, one can consider driving such a transition by going to finite chemical potential for the large $N$ gauge fields at $T=0$, at the cost of introducing Reissner-Nordstr\"{o}m black branes.  At sufficiently large chemical potential, even at zero temperature, the horizon of the extremal Reissner-Nordstr\"{o}m black brane grows large, and the probe branes will transition back to a configuration where they stretch to the horizon instead of reconnecting. It would be interesting to determine the order of this phase transition at zero temperature.}

\item There exist fermionic operators ${\cal O}^{F}_J$ localized at the $J$th lattice site, whose thermal correlation functions in the undimerized phase are known and gapless:
\bea
\label{UVscaling}
\int dt e^{i\omega t}\langle {\cal O}^{F}_J(t) {\cal O}^{F\dagger}_{J'}(0)\rangle&=&i \delta_{J, J'} {\cal G}(\omega),\nonumber\\
{\rm with}\ \ \ {\cal G}(\omega)&\sim&  \omega^{2\Delta - 1} \ \ \ {\rm for}\ \ \  \omega \gg T. \eea

\item In the dimerized phase, the spectrum is gapped and
\be
\lim_{\omega\rightarrow0}\int dt e^{i\omega t}\langle {\cal O}^{F}_J(t) {\cal O}^{F\dagger}_{J'}(0)\rangle=iA_{J, J'}.
\ee
Here, $A_{J, J'}$ is nonzero (generically if and) only if $J=J'$ or $J$ and $J'$ are paired up via dimerization.

\end{enumerate}
For example, for the literal D5 probe theory in ${\rm AdS}_5 \times S^5$, we can take
\begin{equation}
{\cal O}^{F}_J= \chi_{J}^\dagger \lambda_{{\cal N}=4}(J) \chi_J
\end{equation}
and work out $ {\cal G}(\omega)$ and $A_{J, J'}$ as a function of external parameters.
Here $\lambda_{{\cal N}=4}(J)$ is the ${\cal N}=4$ gaugino evaluated at the $J$th lattice site, and $\chi_J$ is the probe fermion associated with the $J$th site. There is also an infinite tower of similar operators of higher conformal dimension. We will, however, keep our discussion abstract.

Note that we are only guaranteed of the scaling form~(\ref{UVscaling}) governed by the (0+1)-dimensional conformal invariance when $\omega \gg T$.\footnote{At very low frequency ${\cal G}(\omega)$ will still approach zero on general grounds, but it may do so with a different scaling dimension $\Delta'$ or in even more complicated ways.}  Therefore, looking forward for
a moment (to the stage where we mix the ${\cal O}^F$s with semi-holographic fermions)
this behavior of the Green's function will be relevant when studying excitations close to the Fermi surface, {\it only if} the disconnected phase persists to very low temperatures
(compared to the Fermi momentum $k_F$).  This is achievable in our models, because the temperature of the dimerization transition is $T_c \sim {1\over a_{\rm defect}}$ \cite{Kachru:2009xf},
where $a_{\rm defect}$ is the lattice spacing for defect fermions, and can be dialed freely; while $k_F\sim\frac{1}{a_{\rm itinerant}}$ is another free parameter, where $a_{\rm itinerant}$ is the lattice spacing for semi-holographic itinerant free fermions, which can also be
adjusted independently. Thus we make a hierarchy $a_{\rm defect}\gg a_{\rm itinerant}$.

\

We now semi-holographically couple this large $N$ field theory to the free band fermion in the spirit of \cite{Faulkner:2010tq}:\footnote{For notational simplicity, we made the free $c$ fermions live on the same lattice sites as the defect fermions do. As just mentioned, however, we should really make the $c$ fermions live on a much finer lattice to get the hierarchy $\frac{1}{a_{\rm itinerant}}\sim k_F \gg T_c\sim\frac{1}{a_{\rm defect}}$. Also as in \S2 of \cite{Faulkner:2010tq}, we have neglected possible spin-orbit effects that could
promote, for example, coupling constants between $c$ and ${\cal O}^F$ in (\ref{coupling}) to be matrices in spin space.}
\bea
 \label{coupling}
 S =&& S_{\rm strong} + \sum_{J, J'} \int dt~\left[ c_{J}^{\dagger} (i\delta_{J,J'}\partial_t + \mu\delta_{J,J'}+ t_{J,J'}) c_{J'} \right]\nonumber\\
 &&+g \sum_{J}  \int dt \left[c_J^\dagger {\cal O}^{F}_J +({\rm Hermitian\ conjugate}) \right].
\eea
Here $t_{J,J'}$ characterizes the band structure of the originally free fermion $c$ sector, which now mixes with the large $N$ dimer model through the coupling constant $g$.

The key insight of \cite{Faulkner:2010tq} is that large $N$ factorization of the field theory (which would work even at small 't Hooft coupling) can be used to infer the
modifications to the two-point functions of the conducting $c$ fermions arising from the coupling $g$.  The $g=0$ Green's function for the $c$
fermions is
 \begin{equation}
\label{GFs}
G_{0}({\bf k},\omega) \equiv -i \frac{1}{N_{\rm l.s.}}\sum_{J,J'} \int dt  e^{i\omega t-i {\bf k}\cdot({\bf x}_J-{\bf x}_{J'})}\langle c_J (t)c_{J'}^{\dagger}(0) \rangle_{g=0}\sim {1\over {\omega - v|{\bf k}-{\bf k}_{F}({\bf k})|}}
\end{equation}
with ${\bf k}_F({\bf k})$ the point on the Fermi surface, closest to the argument ${\bf k}$, and $N_{\rm l.s.}$ the number of lattice sites.
Then we find that for finite coupling $g$, after summing a geometric series of tree-level mixing diagrams,
\begin{equation}
\label{neat}
G_{g}({\bf k},\omega) \sim {1\over {\omega - v|{\bf k}-{\bf k}_{F}({\bf k})| - g^2 {\cal G}({\bf k}, \omega)}}~.
\end{equation}
In particular, for ${\cal G}({\bf k},\omega)=c \omega^{2\Delta -1}$ with $\Delta \leq 1$, one finds a dominant low-frequency behavior characteristic of a non-Fermi liquid which has vanishing quasiparticle residue [with marginal Fermi liquid behavior precisely at
$\Delta = 1$, when the naive $\omega^{2-1}$ is modified to have $\omega {\rm log}(\omega)$ behaviour].     For $\Delta > 1$, the residue does not vanish, but the
theory is still novel in that the
quasiparticle width does not agree with that of standard Fermi liquid theory.
As we described above, these results are true in a regime where $ k_F \gg \omega \gg \frac{1}{a_{\rm defect}}$, where the zero-temperature Green's functions used above should be
a good approximation to the true (finite- but low-temperature) answers.

Now, we are in a position to add one simple observation on top of the basic picture advocated in \cite{Faulkner:2010tq}: in holographic models which undergo a dimerization
transition as in Sec.~\ref{essence}, the phase transition also drives an interesting transition in the structure of the Fermi surface.  The main point is that the low frequency behavior of the Green's function
$\int dt e^{i\omega t}\langle {\cal O}^{F}_J(t) {\cal O}^{F\dagger}_{J'}(0)\rangle$ changes drastically in the dimerization transition. In the undimerized state, we will have non-Fermi liquid behavior just as in \cite{Faulkner:2010tq}. However, in the dimerized phase, the spectrum in the dimer sector is gapped. This means that at low frequencies, instead of exhibiting power-law behavior, $\lim_{\omega\rightarrow0}{\cal G}({\bf k},\omega)=A$ for some nonzero constant $A$. Thus in this phase, we have a conventional Fermi liquid whose Fermi surface is shifted from the original ${\bf k}_{F}$.

Therefore, in this semi-holographic setting, the dimerization transition of Sec.~\ref{essence} becomes a transition between a conventional Fermi liquid phase (dimerized) and a
non-Fermi liquid phase (undimerized).  These transitions are somewhat reminiscent of the phase transitions in Kondo lattice models discussed in \cite{Sachdev:2010um} and references therein.

Finally, we note that if one is purely interested in finding realizations of the non-Fermi liquid phase, without studying phase transitions of the Fermi surface, one can also simply study
the BPS lattice model made only of D5-branes and generalizations thereof.
In this case, the Schwarzschild ${\rm AdS}_5$ black brane (with probe D5-branes wrapping ${\rm AdS}_2$ subspaces in the ${\rm AdS}_5$) correctly captures the physics down to temperatures
of order $T \sim {\lambda^{1/6} \over N^{1/3} a_{\rm defect}}$. After incorporating semi-holographic fermions, now without the constraint $\omega\gg \frac{1}{a_{\rm defect}}$ as there is no dimerization transition, even very low frequency excitations above the Fermi surface are governed by (\ref{UVscaling}) and (\ref{neat}) at large $N$.

\section{Double trace deformation: band structure of dimer vibrations}
\label{band}

\begin{figure}[t]
\includegraphics[scale=0.65,angle=0]{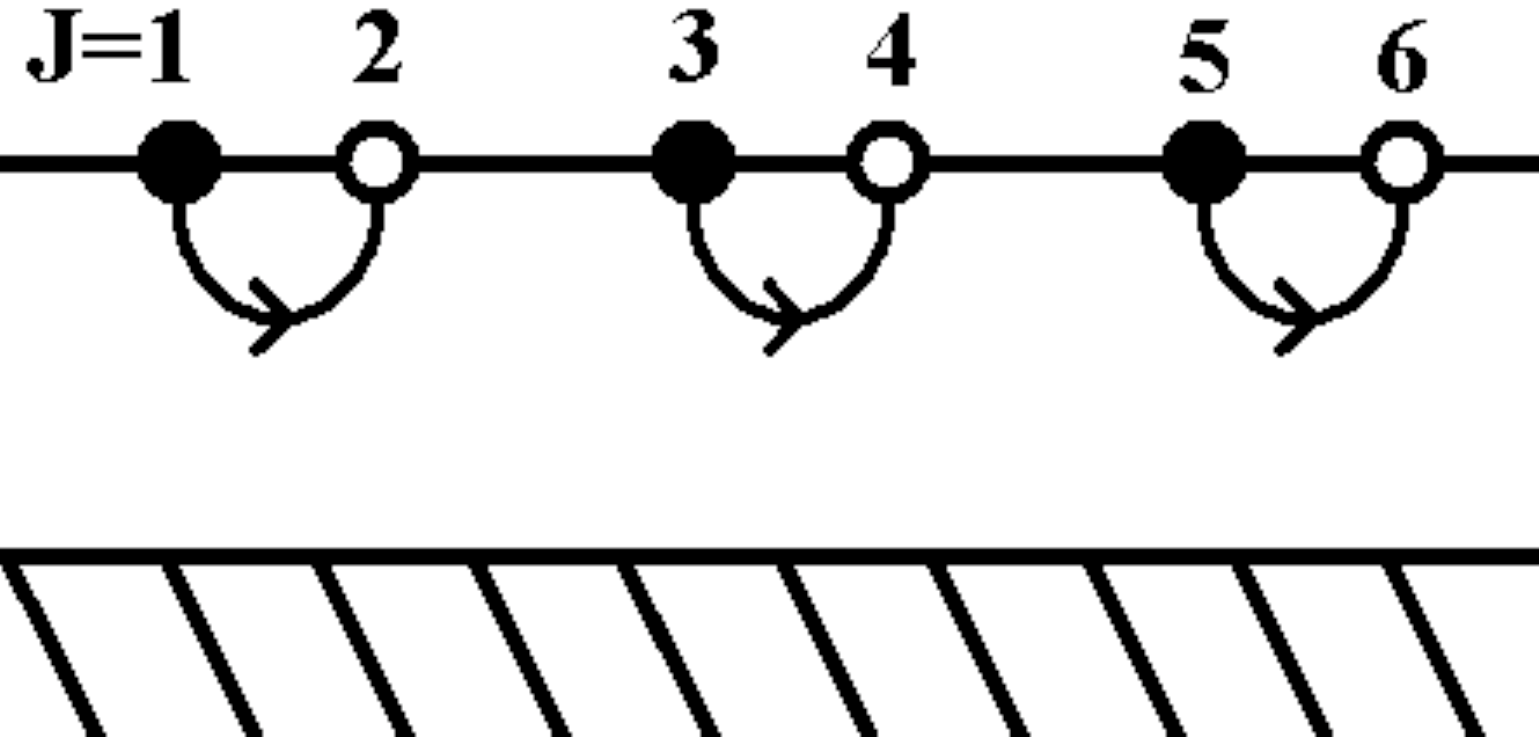}
\caption{Dimerized configuration of interest. Note that we made the distance between $J=(2j+1)$th site and $J=(2j+2)$th site smaller than that between $J=(2j)$th site and $J=(2j+1)$th site, so that we have one unique dimerized configuration below the critical temperature.}
 \label{uniquegrounddimer}
\end{figure}

Next, we visit the landscape of holographic dimer models with certain double trace deformations added to the boundary Lagrangian. For simplicity, let us consider the 1-dimensional array of dimers (see Fig.2). Let us label sites so that the $J=(2j+1)$th site is paired up with the $J=(2j+2)$th site with $j\in {\mathbb Z}$. Our inputs are:
\begin{enumerate}
\item There are bosonic Hermitian operators ${\cal O}^B_J$ which corresponds to a bosonic fluctuation of a probe brane originating from the $J$th site.  The fluctuation take the asymptotic form (for frequency $\omega$)
\be
e^{-i\omega t}\left[\alpha_J(\omega) \left\{ z_J^{1-\Delta}+...\right\}+ \beta_J(\omega) \left\{z_J^{\Delta}+...\right\}\right].
\ee

\item We stay in the dimerized phase where the coefficients $\alpha_{2j+1}(\omega)$, $\beta_{2j+1}(\omega)$, $\alpha_{2j+2}(\omega)$, and $\beta_{2j+2}(\omega)$ are related as follows:
\be
\label{vibe}
\le(\begin{array}{ c c }
\alpha_{2j+2}(\omega) \\
\beta_{2j+2}(\omega) \end{array} \ri)
=
\le(\begin{array}{ c c }
t_{11}(\omega) & t_{12}(\omega) \\
t_{21}(\omega) & t_{22}(\omega)\end{array} \ri)
\le(\begin{array}{ c c }
\alpha_{2j+1}(\omega) \\
\beta_{2j+1}(\omega)\end{array} \ri).
\ee

\end{enumerate}

Originally, the dimers are basically decoupled from each other and each has a discrete vibration spectrum at $\omega=\omega_n$ where $t_{12}(\omega_n)=0$. We now deform the theory by double trace operators, and determine the resulting band structure.

\subsection{Double trace deformation}
We add a double trace deformation of the form
\be
\label{dt}
\Delta L_{\rm d.t.} = h' \sum_{j\in{\mathbb Z}} {\cal O}^B_{2j} {\cal O}^B_{2j+1}
\ee
to the Lagrangian. The effect of double trace deformations on the dual gravitational description is well known \cite{Aharony:2001pa,Witten:2001ua,Berkooz:2002ug,Sever:2002fk,Aharony:2005sh}. In our context, the standard recipe leads to
\be
\label{prop}
\le(\begin{array}{ c c }
\alpha_{2j+1}(\omega) \\
\beta_{2j+1}(\omega) \end{array} \ri)
=\le(\begin{array}{ c c }
0 & h \\
\frac{1}{h} & 0\end{array} \ri)
\le(\begin{array}{ c c }
\alpha_{2j}(\omega) \\
\beta_{2j}(\omega)\end{array} \ri)
\ee
where $h\equiv(2\Delta-1)h'$. Note that this is a relation between $(2j)$th site and $(2j+1)$th site belonging to different dimers, {\it not} a relation between the $(2j+1)$th site and
the $(2j+2)$th site which together form a dimer.

Instead of the totally reflecting boundary conditions implied by the undeformed condition $\alpha_J=0$,
the deformed boundary condition (\ref{prop}) implies that when a fluctuation associated with (the bulk field dual to) ${\cal O}^B_{2j}$ hits the $(2j)$th AdS$_2$ boundary, part of the wave gets reflected back towards the $(2j-1)$th site, but a small fraction (governed by $h$) instead gets transmitted to the $(2j+1)$th site.\footnote{These ``transparent" boundary conditions modify the propagator of the scalar fields with interesting consequences for loop corrections, as
explored in \cite{Porrati:2003sa,Duff:2004wh,Kiritsis:2006hy,Aharony:2006hz}.}

\subsection{Band structure}
In summary, the equation of motion on the probe brane (top-down or bottom-up) gives the relation~(\ref{vibe}) whereas the double trace deformation yields the relation~(\ref{prop}). We now look for most general (spatially normalizable) time-dependent, but nondissipative, modes with these relations. For simplicity, we use the periodic boundary condition with a number of lattice sites $N_{\rm l.s.}=2N_{\rm dimer}$ and then take the thermodynamic limit $N_{\rm dimer}\rightarrow\infty$ at the end.

Starting with a generic $(\alpha_1, \beta_1)$ and evolving through the chain, we get:
\bea
\le(\begin{array}{ c c }
\alpha_2 \\
\beta_2 \end{array} \ri)&=&
\le(\begin{array}{ c c }
t_{11}(\omega) & t_{12}(\omega) \\
t_{21}(\omega) & t_{22}(\omega)\end{array} \ri)
\le(\begin{array}{ c c }
\alpha_1 \\
\beta_1\end{array} \ri) ={\mathbb T}(\omega)\le(\begin{array}{ c c }
\alpha_1 \\
\beta_1\end{array} \ri) ,
\\
\le(\begin{array}{ c c }
\alpha_3 \\
\beta_3 \end{array} \ri)&=&
\le(\begin{array}{ c c }
0 & h \\
\frac{1}{h} & 0\end{array} \ri)
\le(\begin{array}{ c c }
t_{11}(\omega) & t_{12}(\omega) \\
t_{21}(\omega) & t_{22}(\omega)\end{array} \ri)
\le(\begin{array}{ c c }
\alpha_1 \\
\beta_1 \end{array} \ri),...
\eea
Continuing this way, and getting back to the original site, our periodic boundary condition imposes
\be
\le(\begin{array}{ c c }
\alpha_1 \\
\beta_1 \end{array} \ri)=
\le[
\le(\begin{array}{ c c }
h t_{21}(\omega) & h t_{22}(\omega) \\
\frac{t_{11}(\omega)}{h} &\frac{t_{12}(\omega)}{h}\end{array} \ri)
\ri]^{N_{\rm dimer}}
\le(\begin{array}{ c c }
\alpha_1 \\
\beta_1 \end{array} \ri).
\ee
Taking the thermodynamic limit $N_{\rm dimer}\rightarrow\infty$, we conclude that there exists a nondissipative solution with frequency $\omega(k)$ and with Bloch momentum $k$ if and only if
\be
{\rm det}\le[
\le(\begin{array}{ c c }
h t_{21}(\omega(k)) & h t_{22}(\omega(k)) \\
\frac{t_{11}(\omega(k))}{h} &\frac{t_{12}(\omega(k))}{h}\end{array} \ri)
-\le(\begin{array}{ c c }
e^{ika} & 0 \\
0 & e^{ika} \end{array} \ri)
\ri]=0
\ \ \ {\rm with}\ \ k\in\left[-\frac{\pi}{a}, +\frac{\pi}{a}\right]~.
\ee
The band structure is encoded in $\omega(k)$.

\subsection{A simple toy model}
\label{cari}
\begin{figure}[t]
\begin{center}$
\begin{array}{ccc}
\includegraphics[scale=0.5]{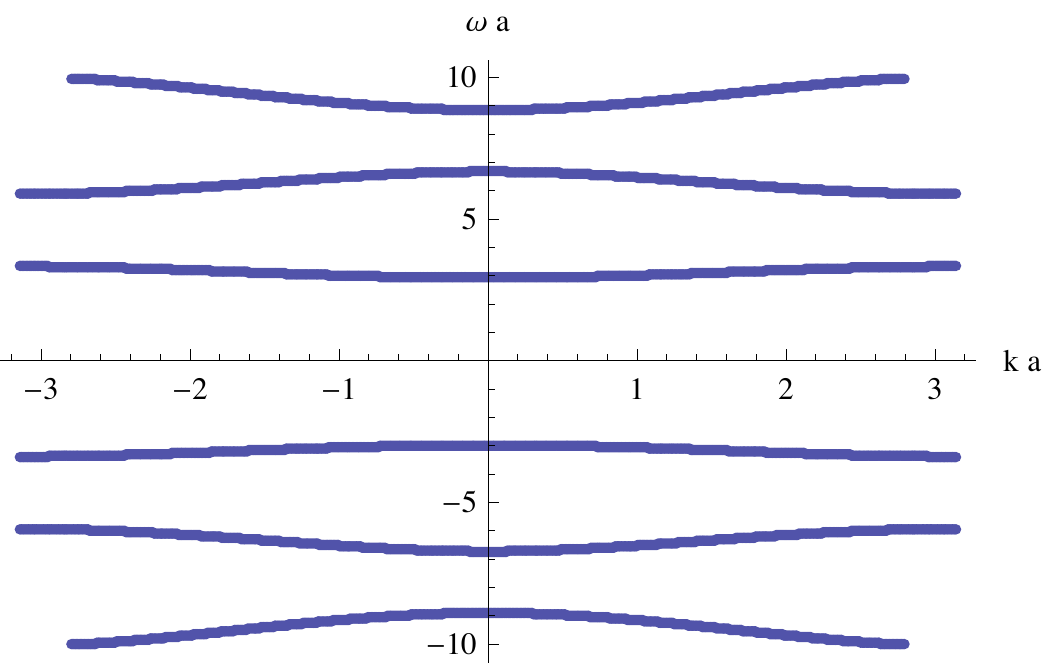}&
\includegraphics[scale=0.5]{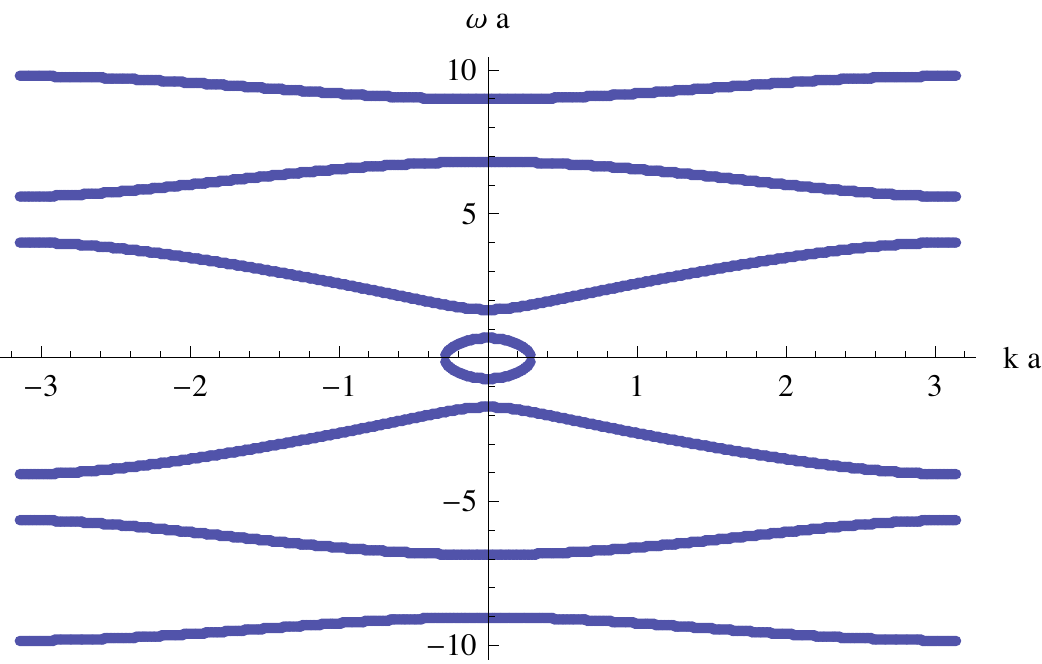}&
\includegraphics[scale=0.5]{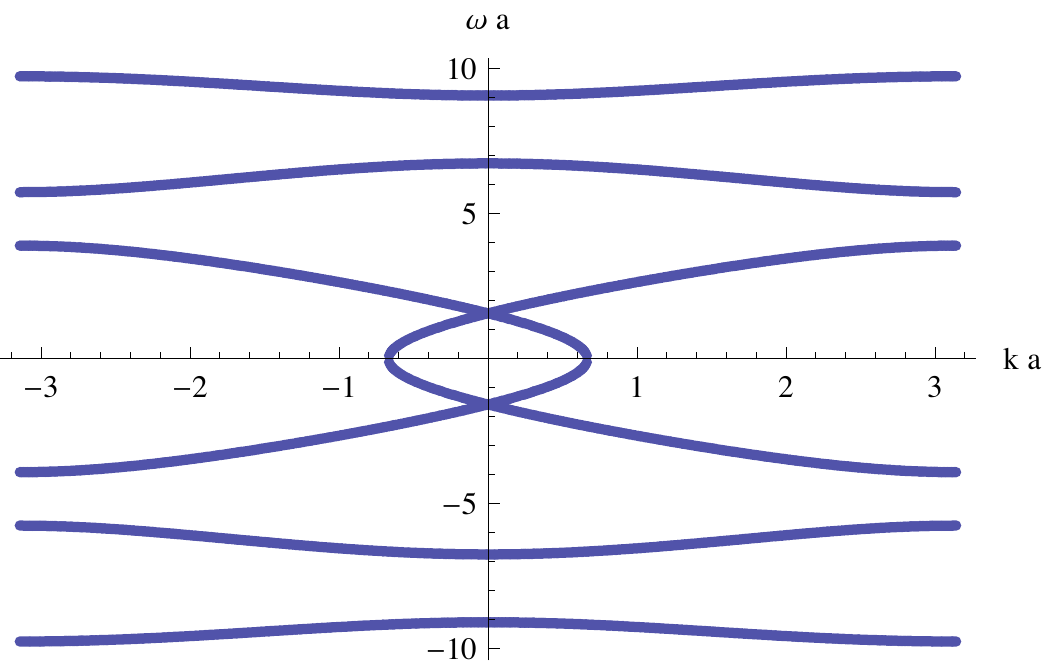}
\end{array}$
\end{center}
\begin{center}$
\begin{array}{cc}
\includegraphics[scale=0.5]{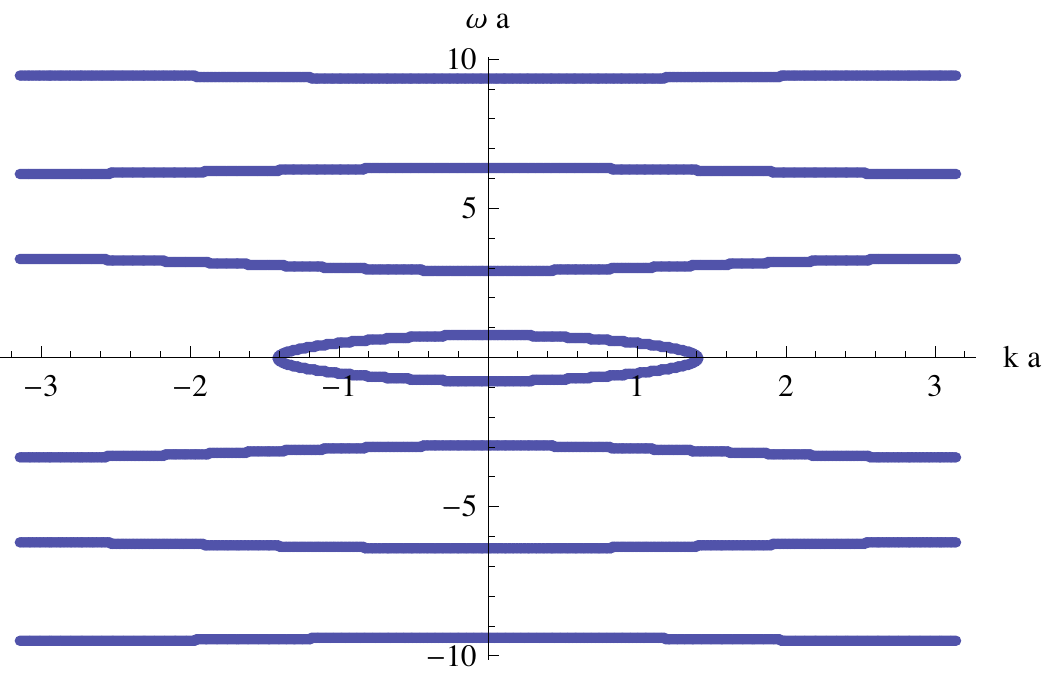}&
\includegraphics[scale=0.5]{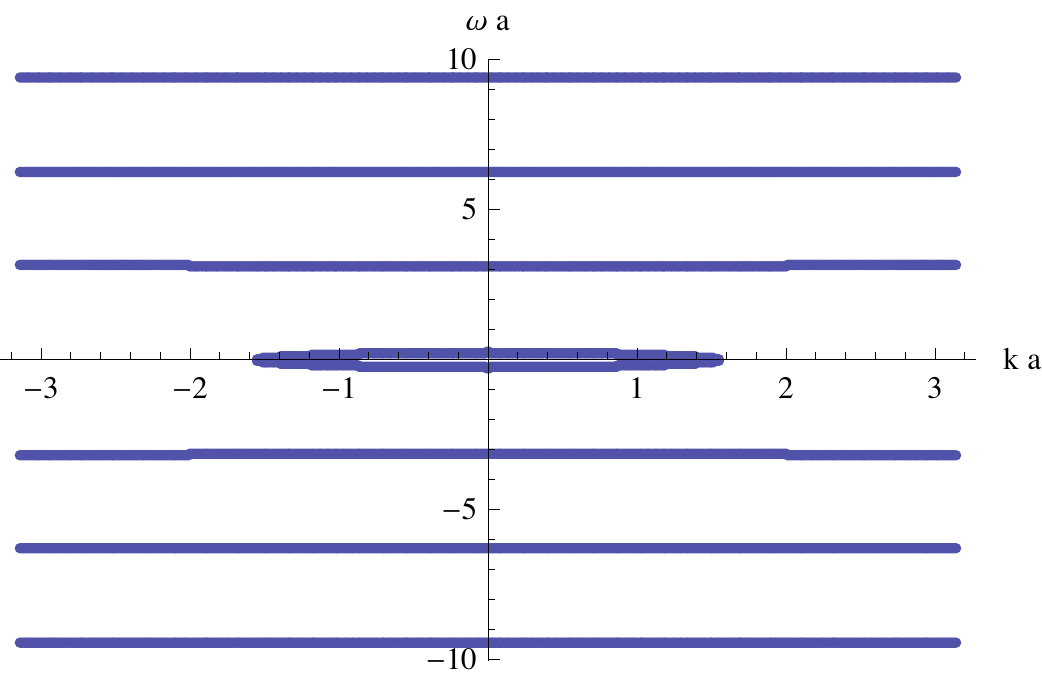}
\end{array}$
\end{center}
\caption{
\label{bands}Band structure of the global AdS$_2$ toy model for $\Delta=1$ and $h=0.05$ (top left panel), $h=0.82$ (top middle panel), $h=1$ (top right panel), $h=5$ (bottom left panel) and $h=50$ (bottom right panel).
}
\end{figure}

If we know
${\mathbb T}(\omega)$,
it is a simple matter to map out the band structure numerically.
As an illustration, let us perform this exercise for the caricature toy model described fully in Appendix~\ref{hyper}.
For the special case of $\Delta=1$ we display the first few bands for several values of $h$ in Fig.~\ref{bands}.\footnote{Note that, as the equations of motion only depend on $\omega^2$, the bands have symmetry around $\omega=0$.} Without the double trace deformation, the full spectrum is just $N_{\rm dimer}$ copies of the spectrum of a single dimer, independent of $k$. We see that for small $h$ we still have a band structure with very narrow bands centered around the mode spectrum of the uncoupled dimers, $\omega_n = \frac{n \pi}{a}$ for nonzero integer $n$. Around $h=0.8$ a new band emerges and this sector becomes gapless, potentially signaling the onset of an instability towards forming a spatially inhomogeneous condensate similar to the one encountered at the Lifshitz point in ferromagnets. As $h$ is the coupling constant of a double trace deformation, such a spatially inhomogeneous condensate would not be visible at the leading order $N$ classical action, but only in the order 1 free energy induced from loops. 
For higher values of $h$ the lowest two bands undergo an interesting reorganization, changing the topology of the band structure. For very large $h$ we once more approach degenerate $k$-independent bands (but now with $\omega\approx 0$ band surviving, as clearly visible in the bottom right panel of Fig.~\ref{bands}), this time with the opposite boundary conditions [in other words, $\alpha_J$ has to vanish at each site for $h=0$ while $\beta_J$ has to vanish at each site for $h=\infty$, giving $t_{21}(\omega_n)=0$ rather than $t_{12}(\omega_n)=0$].\footnote{Incidentally, the analysis of this section goes through in the same way for fermionic operators ${\cal O}^F_J$. It would be interesting to explore implications of this topology change in band structure for such fermionic excitations.}

\section{Towards a Holographic Hubbard Model}
\label{NAhopping}

\begin{figure}[t]
\includegraphics[scale=0.65,angle=0]{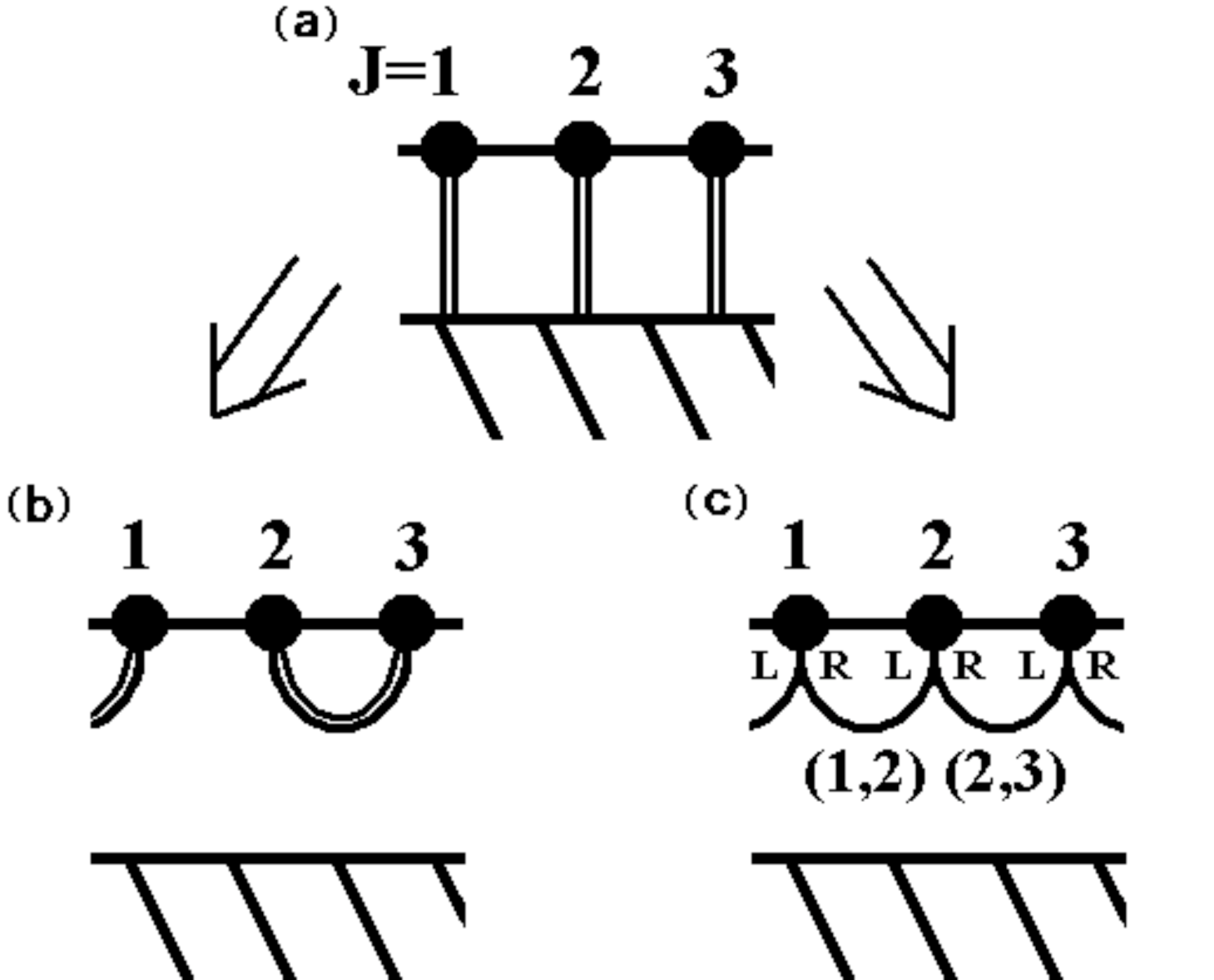}
\caption{(a)The high-temperature phase with two branes at each site, with lattice sites equally spaced.  (b)A possible low-temperature configuration. (c)Another possible low-temperature configuration on which we will focus.}
 \label{Nonabelian}
\end{figure}

While the double trace deformations we introduced allow dimer vibrations, or ``mesons," to propagate through the whole crystal, they do not lead to transport of defect fermions. The double trace deformation we have introduced is the product of two $U(1)_J$ invariant operators, so it preserves all the $U(1)_J$ global symmetries. In particular, the difference in number of defect fermions at the two ends of each dimer, or ``baryon number," is conserved and thus the defect fermions cannot move.

The only way to actually introduce a moving charge carrier seems to be to introduce a charged field or object in the bulk that can effectively carry baryon number.
One potential such object would be the $W$-boson of a $U(2)$ non-Abelian gauge field living on the stacks of branes. So we imagine doing something like:
\begin{enumerate}
\item At each odd $J=(2j+1)$th site we put {\it two} anti-D5-brane, while we put {\it two} D5-brane at each even $J=(2j)$th site, and this time we equally space lattice sites (see Fig.4). At each site, we label two species of defect fermions by $\chi_{L,J}$ and $\chi_{R,J}$.

\item In order to make charge carriers move around, we consider deforming the Lagrangian by a conventional hopping term
\be
\label{hoppingterm}
\Delta L_{\rm hop} = t_h \sum_J \chi^{\dagger}_{L,J} \chi_{R,J} \, + {\rm Hermitian\ conjugate}.
\ee
Here $t_h$ can be a complex coupling constant.
\end{enumerate}
There is a two geometrically distinct configurations with the same energy (see Fig.4). Deformations of the theory as well as $1/N$ corrections can lift this degeneracy. Here we will exclusively focus on the state depicted in Fig.4(c).

The scalar partner of the $W$-boson is dual to an operator of the form $\chi_{L,J}^\dagger \chi_{R,J}$ on a given site, so turning on a nontrivial source for such a field activates a conventional hopping term in the system, rather than the terms quartic in defect fermions we have been implicitly dealing with by resorting to double trace deformations.
This hopping operator affects the dynamics at leading order and is not suppressed like the double trace deformations we considered above. Of course, one can always choose $t_h$ to be a small parameter, and treat the problem perturbatively in $t_h$ for $|t_h| \ll 1$.

In the bulk this operator maps to an off-diagonal component of the $U(2)_J$ current living on the D-branes on the $J$-th site. Note that in the full D3/D5 system the coupled scalar/vector sector on the $J$-th site is dual to the operators $\chi^{\dagger}_{a,J} \chi_{b,J}$, the chiral condensate, and $\chi^{\dagger}_{a,J} \gamma_0 \chi_{b,J}$, the dual current. As $\gamma_0=i$ is just a number, these two operators are really just real and imaginary parts of the hopping term deformation on each site. The labels $a$, $b$ run over $L$ and $R$. There are 4 real operators worth of terms we can add to the Lagrangian, and there are 4 real operators worth of conserved currents dual to the massless gauge field mode in the bulk. As discussed before, in the full D3/D5 system there is a second operator with the same global charge assignments and dimension 3. For minimalistic bottom-up models without such an operator presumably no bulk scalar field mixing with the vector is required.

In order to analyze the effect of turning on the off-diagonal components it is convenient to treat the problem as a simple $U(2)^{N_{\rm l.s.}}$ gauge theory living on a segment of AdS$_2$, where $N_{\rm l.s.}$ denotes the number of lattice sites.
Let us
parametrize the gauge field on each site as
 \be A_{\mu,J}=
\begin{pmatrix}
A^L_{\mu, J}(z,t) & h_{\mu,J}(z,t) \\ h_{\mu,J}^*(z,t) & A^R_{\mu,J}(z,t)
\end{pmatrix}.
\ee
A convenient gauge choice is $A_{z,J}=0$.
For a minimalistic model one can take the intrinsic metric to be a segment of pure AdS$_2$
\be
\label{pureads}
ds^2=\frac{1}{z^2} (- dt^2 + dz^2)
\ee
 running from the UV boundary at $z=0$ to some ``hard wall" at $z=z_*$. The hard wall at $z_*$ simply reflects the fact that we are studying a state of the system in which the branes are reconnected, and do not reach all the way to the horizon.
The detailed state of the field theory is captured by the boundary conditions imposed at the hard wall.

For the purpose of phenomenological model building, the boundary conditions imposed in the IR are part of the input. But there is a particular ``geometric" set of IR boundary conditions that corresponds to the connected bridge configurations we studied before. There, the boundary conditions on the diagonal components of the gauge
fields follow by continuity of the fields and their radial derivatives in the true bridged configuration.  If we let $\sigma$ denote a single
valued radial coordinate along the brane, with $z(\sigma)$ an increasing function on the right brane and a decreasing function on the left brane, then the natural
boundary conditions are:
\be
\label{bcdiagonal}
A_{t,J+1}^L - A_{t,J}^R =0, \,\,\,\,\, \partial_{\sigma} ( A_{t,J+1}^L - A_{t,J}^R)=
\partial_{z}  A_{t,J+1}^L + \partial_z A_{t,J}^R =0 \,\,\,\,\,\, \mbox{at } z=z_*.\ee
Note that these boundary condition gives the desired breaking of the $U(2)^{N_{\rm l.s.}}$ to the $U(1)^{N_{\rm l.s.}}$ associated with the $U(1)$ gauge fields living on the bridges:
 \be
 A_{t,J+1}^L = A_{t,J}^R \equiv A_{t,(J,J+1)}.
 \ee
 Having different boundary conditions on the left and right fields directly breaks each $U(2)$ to $U(1) \times U(1)$ and then, as in the brane setups studied earlier, the boundary conditions ensure that the left $U(1)$ field of the $(J+1)$th site is identified with the right $U(1)$ field of the $J$th site.

Last but not least, we have to determine the boundary conditions on the off-diagonal components of
the gauge field, $h_{t, J}$ and $h_{t, J}^*$.
From the bottom-up point of view the best way to think about the IR boundary conditions is to introduce an ``IR-brane-localized Higgs field." As in the cases of interest the boundary conditions always connect $U(2)$ gauge fields living on neighboring sites, one can introduce complex, bi-fundamental scalars connecting neighboring sites. That is, for every bridge with label $(J,J+1)$, one adds an IR brane localized Lagrangian for a 2 by 2 matrix of scalar fields $\phi_{(J,J+1)}$
\be Tr |D_{\mu} \phi_{(J,J+1)}|^2 = Tr | \partial_{\mu} \phi_{(J,J+1)} - i A_{\mu,J} \phi_{(J,J+1)} + i \phi_{(J,J+1)} A_{\mu,J+1}|^2. \ee
One particularly interesting form of the vacuum expectation value (vev)
of the IR-brane localized Higgs field is
\be \label{vev} \phi_{(J,J+1)} = \begin{pmatrix}  0 & 0 \\ m & 0 \end{pmatrix}.
\ee

 Eq.(\ref{vev}) is the unique choice if we want a vev that only gives quadratic terms mixing $A^R_{t,J}$ with $A^L_{t,J+1}$ but no other quadratic terms involving the diagonal components of the gauge fields. This form of the vev is ``geometric" in the sense that it can describe branes reconnecting as we saw above.
This form of a scalar field expectation value gives rise to IR brane localized mass terms for the gauge fields of the form
\be
\label{irlag}
L_{\rm IR} = \sum_J |m|^2 \left ( (A^L_{t,J+1} - A^R_{t,J})^2 + |h_{t,J}|^2 + |h_{t,J+1}|^2 \right ).
\ee

The boundary conditions on a gauge field with a finite boundary mass matrix $(m^2)^{ab}$ (due to the boundary Higgs) are in general
\cite{Csaki:2003dt,Csaki:2003zu}
\be
\partial_z A_t^a = (m^2)^{ab} A_t^b \,\,\,\,\,\, \mbox{at } z=z_*.
\ee
We see that the above Higgs vev and the resulting IR Lagrangian (\ref{irlag}) in the limit of large $m$ give exactly
the boundary condition (\ref{bcdiagonal}) that we know encode the correct geometric conditions on the bridges, together with a Dirichlet boundary condition on the off-diagonal components of the gauge field.

In this language it is now straightforward to turn on the actual fermion hopping interaction of Eq. (\ref{hoppingterm}). Asymptotically, we have
\be
A_{t,J}^{R,L} = \alpha_J^{R,L} + \beta_J^{R,L}/z, \,\,\,\,\, h_{t, J} = \gamma_J + \delta_J/z.
\ee
Turning on a nontrivial hopping interaction simply tells us that we are studying bulk gauge field
configurations in which we impose the UV boundary condition that $\gamma_J=t_h$.

In the full D3/D5 system it is impossible to study this deformation reliably. For one thing, turning on the off-diagonal gauge field components requires one to know the full non-Abelian DBI action governing the dynamics of these fields. However, this action is not known beyond the few lowest dimension terms in powers of $F_{\mu \nu}$. Even worse, to reliably study the bridged configurations we really need the analogue of the DBI action that governs a brane/anti-brane system including the tachyon field. This is certainly beyond the scope of present-day D-brane technology.

On the other hand, at the level of bottom-up model building, we have assembled all of the ingredients we need to study a holographic realization of a generic lattice model with hopping fermions. Thus, one can take one's favorite unsolved lattice model (for example the Hubbard model), and parametrize one's ignorance by writing down a corresponding higher dimensional brane system with an effective action for both the gauge field and the IR Higgs field.   This action will have free parameters, which should be matched against known properties of the boundary lattice model. While it is not obvious that such a rewriting will be advantageous, it may offer some new approaches to this class of problems, just as bottom-up models of hadron physics have done for the study of QCD.



\bigskip
\centerline{\bf{Acknowledgements}}

We are grateful to J. Polchinski and S. Sachdev for very helpful conversations.  We also thank S. Shenker and E. Silverstein for interesting discussions about related subjects.
S.K. and S.Y. thank the theory group at the University of Washington at Seattle for hospitality during the completion of this work.
S.K. is also happy to acknowledge the warm hospitality of the string theory group at the Kavli Institute for Theoretical Physics and the UCSB Physics Department
while the bulk of this work was completed. He is supported by the NSF under grant no. 0756174, by the DOE under contract DE-AC03-76SF00515, and by the
Stanford Institute for Theoretical Physics.
S.Y. thanks the Kavli Institute for Theoretical Physics for unofficial hospitality and the MIT Center for Theoretical Physics for official hospitality while this work was vaguely in progress. He is supported by the Stanford Institute for Theoretical Physics and NSF Grant No. 0756174.
\appendix

\section{Hypergeometric-ology}
\label{hyper}
In this section we present a simple toy model of probe brane fluctuations in the dimerized phase. On the probe brane, let us suppose that there is a scalar field governed by the following effective action:
\be
S_{\rm toy}=\int dtdx\sqrt{-g}\le[-g^{\mu\nu}(\partial_{\mu}\phi^{*})(\partial_{\nu}\phi)-m^2|\phi|^2\ri],
\ee
where $g_{\mu\nu}$ is a caricature ``two-AdS$_2$" induced metric given by
\be
g_{\mu\nu}^{({\rm caricature})} dx^{\mu} dx^{\nu}=\frac{1}{\le[{\rm cos}\le\{\pi\le(\frac{x}{a}-\frac{1}{2}\ri)\ri\}\ri]^2}\le(-dt^2+dx^2\ri),\ \ \ x\in[0, a].
\ee
This describes a global AdS$_2$ on the bridge.

Making a coordinate transformation to $\rho\equiv \pi(\frac{x}{a}-\frac{1}{2})$, we get:
\be
\le[-\partial_{\rho}^2+\frac{(\frac{am}{\pi})^2}{{\rm cos}^2 \rho}\ri]\phi_{\omega}(\rho)=\le(\frac{a\omega}{\pi}\ri)^2\phi_{\omega}(\rho),\ \ \ \rho\in\left[-\frac{\pi}{2}, +\frac{\pi}{2}\right].
\ee
By a further transformation
\bea
v&\equiv&\frac{1+{\rm sin}\rho}{2},\ \ \ v\in[0,1],\\
\phi_{\omega}(v)&\equiv&\le(\frac{1}{v(1-v)}\ri)^{\frac{1}{4}}\psi_{\omega}(v),
\eea
the equation of motion can be brought into hypergeometric form:
\be
\frac{\partial^2\psi_{\omega}}{\partial v^2}+\frac{1}{4v^2(1-v)^2}\le[\le\{1-4\le(\frac{a\omega}{\pi}\ri)^2\ri\}v^2-\le\{1-4\le(\frac{a\omega}{\pi}\ri)^2\ri\}v+\le\{\frac{3}{4}-\le(\frac{am}{\pi}\ri)\ri\}\ri]\psi_{\omega}=0.
\ee
We can now bring this equation into the standard Gauss' hypergeometric form:
\bea
a_{\rm hyper}&\equiv&\frac{a\omega}{\pi}+\frac{1}{2},\ \ \ b_{\rm hyper}\equiv-\frac{a\omega}{\pi}+\frac{1}{2},\ \ \ c_{\rm hyper}(c_{\rm hyper}-2)\equiv\le(\frac{am}{\pi}\ri)^2-\frac{3}{4}\\
\psi_{\omega}(v)&\equiv&\le(\frac{1}{v}\ri)^{\frac{-c_{\rm hyper}}{2}}\le(\frac{1}{1-v}\ri)^{\frac{c_{\rm hyper}-a_{\rm hyper}-b_{\rm hyper}-1}{2}}F_{\omega}(v),\\
0&=&\le[v(1-v)\frac{d^2}{dv^2}+(c_{\rm hyper}-(a_{\rm hyper}+b_{\rm hyper}+1)v)\frac{d}{dv}-a_{\rm hyper}b_{\rm hyper}\ri]F_{\omega}(v).
\eea
After reorganizing a bit we get\footnote{Note that $v\sim\le(\rho+\frac{\pi}{2}\ri)^2$ as $\rho\rightarrow-\frac{\pi}{2}$ whereas $(1-v)\sim\le(\rho-\frac{\pi}{2}\ri)^2$ as $\rho\rightarrow\frac{\pi}{2}$, giving rise to $\frac{1}{2}$ in the exponents.}
\bea
\nonumber
\Delta(\Delta-1)&\equiv&\le(\frac{ma}{\pi}\ri)^2 \ \ \ [{\rm choose\ the\ positive\ root\ so\ that}\  (\Delta-1)\geq -\Delta].\\
\phi(v)&=&\le(\frac{1}{v}\ri)^{-\frac{\Delta}{2}}\le(\frac{1}{1-v}\ri)^{\frac{\Delta-1}{2}}[\alpha(\omega)\le(\frac{1}{v}\ri)^{\frac{2\Delta-1}{2}}\le\{_2F_1\le(1-\Delta+\le(\frac{a\omega}{\pi}\ri), 1-\Delta-\le(\frac{a\omega}{\pi}\ri); \frac{3}{2}-\Delta; v\ri)\ri\}\nonumber\\
&&\nonumber \ \ \ \ \ \ \ \ \ \ \ \ \ \ \ \ \ \ +\beta(\omega)\le\{_2F_1\le(\frac{1}{2}+\le(\frac{a\omega}{\pi}\ri), \frac{1}{2}-\le(\frac{a\omega}{\pi}\ri); \frac{1}{2}+\Delta; v\ri)\ri\}]\\
&=&\le(\frac{1}{v}\ri)^{-\frac{\Delta}{2}}\le(\frac{1}{1-v}\ri)^{\frac{\Delta-1}{2}}[\alpha'(\omega)\le\{_2F_1\le(\frac{1}{2}+\le(\frac{a\omega}{\pi}\ri), \frac{1}{2}-\le(\frac{a\omega}{\pi}\ri); \frac{3}{2}-\Delta; 1-v\ri)\ri\}\nonumber\\
&&\nonumber \ \ \ \ \ \ \ \ \ \ \ \ \ \ \ \ \ \ +\beta'(\omega)\le(\frac{1}{1-v}\ri)^{\frac{1-2\Delta}{2}}\le\{_2F_1\le(\Delta+\le(\frac{a\omega}{\pi}\ri), \Delta-\le(\frac{a\omega}{\pi}\ri); \frac{1}{2}+\Delta; 1-v\ri)\ri\}].
\eea

The hypergeometric functions appearing above are connected through frequency-dependent matrices:\footnote{They are also $\Delta$-dependent, where $\Delta$ is a parameter of the theory.}
\bea
&&\le(\begin{array}{ c c}
_2F_1\le(\frac{1}{2}+\le(\frac{a\omega}{\pi}\ri), \frac{1}{2}-\le(\frac{a\omega}{\pi}\ri); \frac{1}{2}+\Delta; v\ri)\\
\le(\frac{1}{v}\ri)^{\frac{2\Delta-1}{2}}\ _2F_1\le(1-\Delta+\le(\frac{a\omega}{\pi}\ri), 1-\Delta-\le(\frac{a\omega}{\pi}\ri); \frac{3}{2}-\Delta; v\ri) \end{array} \ri) \nonumber \\
\nonumber
&&=
\le(\begin{array}{ c c }
t_{12}(\omega)& t_{22}(\omega) \\
t_{11}(\omega) & t_{21}(\omega)\end{array} \ri)
\le(\begin{array}{ c c}
_2F_1\le(\frac{1}{2}+\le(\frac{a\omega}{\pi}\ri), \frac{1}{2}-\le(\frac{a\omega}{\pi}\ri); \frac{3}{2}-\Delta; 1-v\ri)\\
\le(\frac{1}{1-v}\ri)^{\frac{1-2\Delta}{2}}\ _2F_1\le(\Delta+\le(\frac{a\omega}{\pi}\ri), \Delta-\le(\frac{a\omega}{\pi}\ri); \frac{1}{2}+\Delta; 1-v\ri) \end{array} \ri),
\eea
where $t_{ij}(\omega)$ can be expressed in terms of Gamma functions as
\bea
t_{11}(\omega)&=&\frac{\Gamma\le(\frac{3}{2}-\Delta\ri)\Gamma\le(\Delta-\frac{1}{2}\ri)}{\Gamma\le(\frac{1}{2}+\frac{a\omega}{\pi}\ri)\Gamma\le(\frac{1}{2}-\frac{a\omega}{\pi}\ri)},\\
t_{12}(\omega)&=&\frac{\Gamma\le(\Delta+\frac{1}{2}\ri)\Gamma\le(\Delta-\frac{1}{2}\ri)}{\Gamma\le(\Delta+\frac{a\omega}{\pi}\ri)\Gamma\le(\Delta-\frac{a\omega}{\pi}\ri)},\\
t_{21}(\omega)&=&\frac{\Gamma\le(\frac{3}{2}-\Delta\ri)\Gamma\le(\frac{1}{2}-\Delta\ri)}{\Gamma\le(1-\Delta+\frac{a\omega}{\pi}\ri)\Gamma\le(1-\Delta-\frac{a\omega}{\pi}\ri)},\\
{\rm and}\ \ \ t_{22}(\omega)&=&\frac{\Gamma\le(\Delta+\frac{1}{2}\ri)\Gamma\le(\frac{1}{2}-\Delta\ri)}{\Gamma\le(\frac{1}{2}+\frac{a\omega}{\pi}\ri)\Gamma\le(\frac{1}{2}-\frac{a\omega}{\pi}\ri)}.
\eea

\bibliography{bloch}
\bibliographystyle{apsrev}

\end{document}